\documentclass[a4paper,twocolumn,11pt,accepted=2024-11-26]{quantumarticle}
\pdfoutput=1
\usepackage[utf8]{inputenc}
\usepackage[T1]{fontenc}
\usepackage{amsmath}
\usepackage{xurl}
\usepackage{hyperref}
\hypersetup{breaklinks=true}

\usepackage{tikz}
\usepackage{lipsum}

\usepackage{amsmath,amsfonts,amssymb,amsthm,amsbsy,mathtools}
\usepackage{graphicx}
\usepackage{dcolumn}
\usepackage{bm}
\usepackage{physics} 
\usepackage{bbold}
\usepackage{mathtools}
\usepackage{xcolor}
\usepackage{tikz}
\usepackage{tikz-cd}
\usepackage{comment}
\usepackage{subfigure}
\usepackage{soul}
\usepackage{framed}
\usepackage{mdframed}
\usepackage{csquotes}
\usepackage{appendix}
\usepackage[numbers,sort&compress]{natbib}
\usepackage[normalem]{ulem}
\usepackage{microtype}
\usepackage[english]{babel}
\usepackage{hyphenat}

\newcommand{\F}{\mathcal F}

\newcommand\vk{\textcolor{black}}
\newcommand\cb{\textcolor{black}}

\title{Quantum conformal symmetries for spacetimes in superposi\-tion}

\author{Viktoria Kabel}
\thanks{These two authors contributed equally to this work.}
\affiliation{University of Vienna, Faculty of Physics, Vienna Doctoral School in Physics, and  Vienna Center for Quantum Science and Technology (VCQ) , Boltzmanngasse 5, A-1090 Vienna, Austria}
\affiliation{Institute for Quantum Optics and Quantum Information (IQOQI),
Austrian Academy of Sciences, Boltzmanngasse 3, A-1090 Vienna, Austria}
\orcid{0000-0002-0776-3612}

\author{Anne-Catherine de la Hamette}
\thanks{These two authors contributed equally to this work.}
\affiliation{University of Vienna, Faculty of Physics, Vienna Doctoral School in Physics, and  Vienna Center for Quantum Science and Technology (VCQ) , Boltzmanngasse 5, A-1090 Vienna, Austria}
\affiliation{Institute for Quantum Optics and Quantum Information (IQOQI),
Austrian Academy of Sciences, Boltzmanngasse 3, A-1090 Vienna, Austria}
\orcid{0000-0003-4811-822X}

\author{Esteban Castro-Ruiz}
\affiliation{Institute for Theoretical Physics, ETH Zurich,  Switzerland}

\author{\v{C}aslav Brukner}
\affiliation{University of Vienna, Faculty of Physics, Vienna Doctoral School in Physics, and  Vienna Center for Quantum Science and Technology (VCQ) , Boltzmanngasse 5, A-1090 Vienna, Austria}
\affiliation{Institute for Quantum Optics and Quantum Information (IQOQI),
Austrian Academy of Sciences, Boltzmanngasse 3, A-1090 Vienna, Austria}

\begin{document}

\maketitle
\begin{abstract}
  Without a complete theory of quantum gravity, the question of how quantum fields and quantum particles behave in a superposition of spacetimes seems beyond the reach of theoretical and experimental investigations. Here we use an extension of the quantum reference frame formalism to address this question for the Klein-Gordon field residing on a superposition of conformally equivalent metrics. Based on the group structure of ``quantum conformal transformations'', we construct an explicit quantum operator that can map states describing a quantum field on a superposition of spacetimes to states representing a quantum field with a superposition of masses on a Minkowski background. This constitutes an extended symmetry principle, namely invariance under quantum conformal transformations. The latter allows to build an understanding of superpositions of diffeomorphically non-equivalent spacetimes by relating them to a more intuitive superposition of quantum fields on curved spacetime. Furthermore, it can be used to import the phenomenon of particle production in curved spacetime to its conformally equivalent counterpart, thus revealing new features in Minkowski spacetime with modified Klein-Gordon mass.
\end{abstract}

\section{Introduction} \label{sec: introduction}

Imagine you are living in a universe expanding in a quantum superposition of different Hubble rates. How would fields and particles behave in such a world? Since we lack a complete and generally accepted theory of quantum gravity, an answer to this question remains currently out of reach. Here, we employ an extension of the quantum reference frame formalism 
to relate this exotic situation to a corresponding one that is more easily captured by quantum field theory on curved spacetime (QFTCS). 
More precisely, we map a massive Klein-Gordon (KG) field that resides on a superposition of conformally related spacetimes to a KG field with a superposition of mass parameters on a fixed spacetime. In that way, we establish an equivalence between a situation with a superposition of spacetimes and one with a definite, classical spacetime. Doing so, we identify an additional symmetry of the KG equation -- a \emph{quantum conformal symmetry} -- which goes beyond its invariance under diffeomorphisms.

This additional symmetry of the KG equation enables us to view the two physical situations as equivalent in the present context and can thus serve as a heuristic principle that can tell us about the symmetries of a potential theory of quantum gravity. Such a strategy can be embedded in a larger research program aimed at utilizing symmetries to facilitate the construction of a full theory of quantum gravity, inspired by the historical development of general relativity and quantum field theory \cite{hardy2019}. Recent works have proposed concrete extended symmetry principles, leading to, for example, the quantum relativity of superpositions \cite{zych_relativity_2018}, an equivalence principle for superpositions of spacetimes \cite{giacomini2021einsteins,giacomini2021quantum}, or an extended principle of covariance \cite{delaHamette2021falling}. These symmetry principles are implemented within the framework of quantum reference frame (QRF) transformations. Initially seen as changes between the perspectives of quantum systems \cite{Angelo_2011, Angelo_2012,Giacomini_2019, vanrietvelde2018change, Giacomini_spin, castroruiz2019time, hoehn2019trinity, delaHamette2020, Krumm_2021, mikusch2021transformation, castroruiz2021relative, delahamette2021perspectiveneutral, delahamette2021entanglementasymmetry, Cepollaro_2021}, the interpretation of QRF changes has broadened to encompass a more abstract understanding in the sense of changes of quantum coordinate systems  (e.g.~\cite{hardy2019, giacomini2021einsteins, Giacomini_2021, delahamette2021perspectiveneutral, delaHamette2021falling}). The present work takes this abstraction further by going beyond quantum coordinate transformations and promotes conformal symmetries to the quantum level. In particular, we are able to identify the group structure underlying the conformal transformations of the modified KG equation, which allows us to directly quantize the transformations using the methods of Ref.~\cite{delaHamette2020}. The resulting QRF change operator permits us to map between arbitrary quantum states describing a superposition of semiclassical metrics and the quantum KG field. It thus goes beyond the purely formal level by providing an \emph{explicit} method to relate concrete physical situations. It further goes beyond previous work in that it allows us to treat superpositions of non-diffeomorphically related metrics on a global level. Ref.~\cite{Foo_2020} already treated superpositions of de Sitter spacetimes by introducing an additional control degree of freedom. Note, however, that a superposition of non-diffeomorphic metrics is not required to speak of
\enquote{genuine quantum superpositions of spacetimes} \cite{foo2022}: despite other claims the case treated in Ref.~\cite{delaHamette2021falling} already involved genuine superpositions. Even though Ref.~\cite{delaHamette2021falling} considers diffeomorphic metrics in superposition, the presence of additional systems such as the probe system (and additional fields in general) breaks the symmetry and leads to non-diffeomorphic \emph{physical situations}. Furthermore, Ref.~\cite{giacomini2021einsteins} also treats superpositions of non-diffeomorphically related metrics and shows that they can be locally transformed into Minkowski spacetime, demonstrating the quantum Einstein equivalence principle.

In summary, the present work provides the following three advancements: First, we further abstract the notion of QRFs, focusing on the property that they give rise to a factorization of the total Hilbert space in which the state of the system, which is identified with the frame, factorizes out \cite{delaHamette2020}. Secondly, we analyze superpositions of a set of metrics that are not related by diffeomorphisms. In doing so, we follow \cite{Foo_2020, giacomini2021einsteins, Giacomini_2021, foo_2021} but go beyond these works by introducing a \emph{concrete} QRF transformation operator that renders the spacetime \emph{globally} definite. Furthermore, we go beyond our recent work \cite{delaHamette2021falling} since the transformations established therein allow to make the metric globally definite but are only applicable to spacetimes sourced by superpositions of mass configurations related by Euclidean symmetries. Finally, as we will see below, can consider for the first time \emph{quantum} fields as probes of spacetimes in superposition.

Besides the formal development of quantum conformal transformations, the identified symmetry allows for a better understanding of both the superposition of spacetimes in the presence of a KG field as well as the corresponding modified Minkowski spacetime. On the one hand, in relating the former to a definite spacetime inhabited by quantum fields in superposition of different masses, it sharpens our intuition for the meaning of curved spacetimes in superposition. On the other hand, we use methods of QFTCS to identify the phenomenon of particle production in Minkowski spacetime with a modified mass in superposition.

In Sec.~\ref{sec: Symmetries Conformal}, we begin by reviewing the symmetry properties of the KG equation in curved spacetime. We identify the underlying symmetry group and show that the symmetry transformations form a representation thereof. In a concrete example, we illustrate the transformations relating the metric, the field, and the KG mass on an FLRW spacetime to their counterparts in Minkowski spacetime. In Sec.~\ref{sec: quantum case}, we take the analysis to the quantum level by considering quantized KG fields on a superposition of semiclassical curved spacetimes. We provide the explicit quantum operator that maps between physical situations related by the quantum extension of the aforementioned symmetries. Finally, we use this operator to import results from QFTCS to derive particle production in Minkowski spacetime with a modified mass term in superposition in Sec.~\ref{sec: particle production}. We summarize our findings in Sec.~\ref{sec: conclusion}, relate them to other existing work, and give an outlook on future research directions.

\section{Conformal invariance: Symmetries of the KG Field} \label{sec: Symmetries Conformal}
Let us start by reviewing the symmetry properties of the Klein-Gordon (KG) equation in curved spacetime,
\begin{align}
    \square_g \phi(x) \equiv g^{\mu\nu}\nabla_\mu\nabla_\nu\phi(x) =  m^2\phi(x). \label{eq: KG curvedST}
\end{align}
While the standard KG equation is not invariant under conformal transformations, it has been shown \cite{Wald_gr} that by adding a non-minimal coupling term, the new equation
\begin{align}
    (\square_g +\xi R)\phi(x) = m^2\phi(x)
\end{align}
is invariant under $g_{ab}\to \Omega^2(x) g_{ab}$ and $\phi\to \Omega^{-1}(x)\phi$ only for $\xi = \frac{1}{6}$ and $m=0$.\footnote{Here, we use abstract index notation for tensors, which indicates the type of object in a coordinate-free way rather than its components in any particular basis.} The coupling term $\xi R$ can also be interpreted as contributing to the mass of the KG field.\footnote{A justification for this step can be seen by inspecting the action for the KG field in a non-dynamical spacetime: $\mathcal{S} = \int_Md^4x \sqrt{-g}\left( \frac{1}{2}g^{\mu\nu}\partial_\mu \phi \partial_\nu \phi - \xi R\phi^2 + \frac{1}{2}m^2\phi^2\right)$. Since both the $\xi R$ and the mass term are proportional to $\phi^2$, they can be combined in a single term that can be interpreted as a modified mass term for the KG field. That is, we can write $\mathcal{S} = \int_Md^4x \sqrt{-g}\left( \frac{1}{2}g^{\mu\nu}\partial_\mu \phi \partial_\nu \phi + \frac{1}{2}(m^2-2\xi R)\phi^2\right)$. This leads us to introduce the spacetime-dependent mass term $M^2(x) = m^2 - 2\xi R(x)$.}
If we take this interpretation, we can retain the conformal invariance even for the minimally coupled KG equation with non-vanishing mass. This, however, requires introducing a spacetime-dependent transformation of the mass. For those who consider an evolving mass parameter to be too peculiar, note that it can just as well be seen as an additional potential for the KG field (see e.g.~the discussion around Eq.~(4.1) in Ref.~\cite{Agullo_2013}).

Formally, the overall transformation of the fields, including the mass term, can be expressed as follows. Consider two conformally related metrics $g_1$ and $g_2$ with $(g_2)_{ab}=\Omega^2(x)(g_1)_{ab}$ and a KG field with mass $m_1$ on spacetime $g_1$, which satisfies the KG equation
\begin{align}
    \square_{g_1} \phi_1(x) = m_1^2 \phi_1(x). \label{eq: KG g1 metric}
\end{align}
Next, following  Wald [\cite{Wald_gr}, pp. 445-446], perform the transformation 
\begin{align}
&\begin{pmatrix}
     (g_1)_{ab} \\ \phi_1(x) \\ m_{1}^2
    \end{pmatrix}
    \to \mathcal{F}(\Omega)\left[  \begin{pmatrix}
     (g_1)_{ab} \\ \phi_1(x) \\ m_{1}^2
    \end{pmatrix}\right]\equiv \begin{pmatrix}
     (g_2)_{ab} \\ \phi_2(x) \\ m_{2}^2(x)
    \end{pmatrix} \nonumber\\ =
    &\resizebox{0.9\columnwidth}{!}{$\begin{pmatrix}
     \Omega^2(x) (g_1)_{ab} \\ \Omega^{-1}(x)\phi_1(x) \\ \frac{1}{\Omega^2(x)}\left[ m_1^2-(g_1)^{\mu\nu}(x) \left( \frac{\partial_\mu \partial_\nu \Omega(x)}{\Omega(x)} - (\Gamma^{(1)})_{\mu\nu}^\rho \frac{\partial_\rho \Omega(x)}{\Omega(x)} \right) \right]
    \end{pmatrix}.$}  \label{eq: summary transformations conf}
\end{align}
Then, one can show that $\phi_2(x)$ satisfies the KG equation
\begin{align}
    \square_{g_2} \phi_2(x) = m_2^2(x) \phi_2(x) \label{eq: KG g2 metric}
\end{align} 
for $g_2$ and $m_2^2(x)$. Note that transformation \eqref{eq: summary transformations conf} can in principle map to pathological situations, that is, the mass term $m_2^2(x)$ of the KG field can become negative. In order to retain the interpretation of $m_2^2(x)$ as the square of a mass term, it is thus important to restrict the codomain of this transformation to non-negative mass parameters. As a consequence, there exist cases where $(g_1)_{ab}$ and $(g_2)_{ab}$ are conformally related and yet transformation \eqref{eq: summary transformations conf} does not describe a map to a situation with a straightforward physical interpretation. While this restriction to the codomain of non-negative mass parameters does limit the general applicability of transformation \eqref{eq: summary transformations conf}, it also makes precise in which cases a superposition of spacetimes can be transformed into a superposition of mass parameters and in which it cannot, once we consider the extension to the quantum level.\\

In light of the goal of this work, which is to generalize the conformal invariance of the KG equation to the quantum level, it is important that the transformation \eqref{eq: summary transformations conf} forms a representation of a particular symmetry group. We show in App.~\ref{app: invariance conformal} that this holds true for the symmetry group $(C^2(\mathbb{R}^{3+1},\cdot))$. Specifically, the composition rule is given by $\mathcal{F}_X(\Omega_1(x))\cdot\mathcal{F}_X(\Omega_2(x))=\mathcal{F}_X(\Omega_1(x)\cdot \Omega_2(x))$ for $X=g,\phi,m$.\\

Physically relevant examples for such conformally related spacetimes are all FLRW metrics for arbitrary spacetime-dependent scale factor $a(x)$ and spatial curvature, including anti-de Sitter space \cite{classification_97}. To illustrate the above symmetry, let us explore it in the concrete case of a spatially flat FLRW spacetime with constant Hubble parameter $H \equiv \frac{\dot{a}}{a}$. That is, consider the line element
\begin{align}
   ds^2 = -dt^2 + e^{2Ht}\delta_{ij}dx^i dx^j.
\end{align}

Given two different FLRW metrics characterized by constants $H_1$ and $H_2=H_1+h$ respectively, the transformation \eqref{eq: summary transformations conf} takes on the form
\begin{align}
\begin{pmatrix}
     (g_1)_{ab} \\ \phi_1(t,x) \\ m_{1}^2
    \end{pmatrix}
    &\to \mathcal{F}(h)\left[  \begin{pmatrix}
     (g_1)_{ab} \\ \phi_1(t,x) \\ m_{1}^2
    \end{pmatrix} \right] \equiv \begin{pmatrix}
     (g_2)_{ab} \\ \phi_2(t,x) \\ m_{2}^2(t)
    \end{pmatrix}\nonumber\\
    = &\begin{pmatrix}
     e^{2ht}(g_1)_{ab} \\ e^{-ht}\phi_1(t,x) \\ e^{-2ht}(m_1^2-2H_1^2)+2(H_1+h)^2
    \end{pmatrix}. \label{eq: transformations FLRW}
\end{align}
It is easy to see that $\F$ forms a representation of the underlying symmetry group  $(\mathbb{R},+)$ since $\F_X(h=0)=\mathrm{Id}$, $\F_X(-h)=(\F_X(h))^{-1}$, and the composition rule 
\begin{align}
    \mathcal{F}_X(h_2)\mathcal{F}_X(h_1) = \F_X(h_1+h_2)
\end{align}
is satisfied for all $X=g,\phi,m$. Here, $\mathcal{F}_X(h)$ denotes the action of $\mathcal{F}(h)$ on the metric, the field, and the mass separately.\\ 

In particular, the transformation \eqref{eq: transformations FLRW} allows us to map a situation in such an FLRW spacetime $g_{ab}$ to flat Minkowski spacetime $\eta_{ab}$:
\begin{align}
\begin{cases}
      g_{ab} &\to \eta_{ab}=e^{-2Ht}g_{ab}, \\
    \phi &\to \tilde{\phi} \equiv e^{Ht} \phi, \\
    m^2 &\to M^2(t) \equiv e^{2Ht}(m^2 - 2H^2).\label{eq: FLRWtoFlat}
\end{cases}
\end{align}
This means that the field $\phi$ satisfies the KG equation on $g_{ab}$ with mass term $m^2$ if and only if $\tilde{\phi}$ is a solution to the KG equation on $\eta_{ab}$ with the time-dependent effective mass term  $M^2(t)$.
This can be seen straightforwardly by inserting the expressions \eqref{eq: FLRWtoFlat} into the KG equation on Minkowski spacetime (see App.~\ref{app: invariance FLRW}).

\begin{figure*}
    \centering
    \subfigure[A KG field with fixed mass on a conformally flat spacetime.]
    {
        \centering
        \hfill
        \includegraphics[scale=0.23]{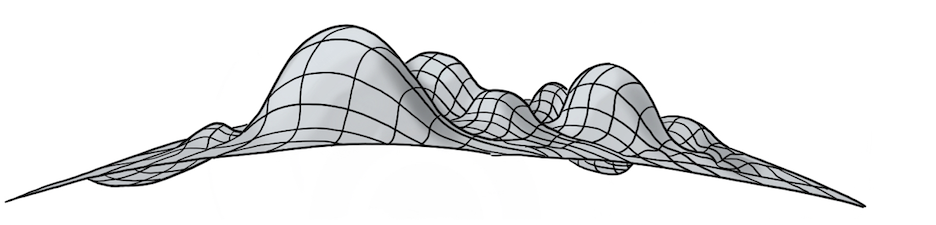}
        \label{fig: leftmanifolds}
    }
    \qquad
    \subfigure[A KG field with time-dependent mass term on flat spacetime.]
    {
        \centering
        \hfill
        \includegraphics[scale=0.23]{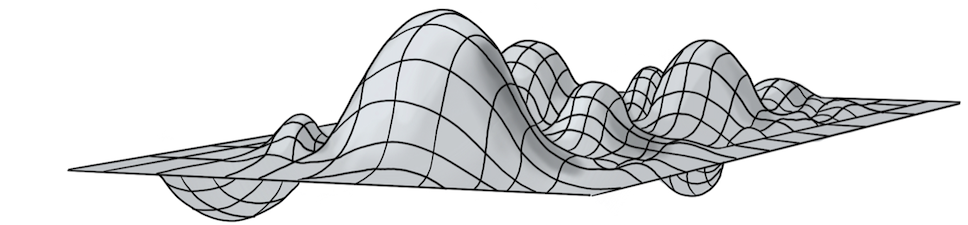}
         \label{fig: rightmanifolds}
    }
    \caption{A schematic pictorial illustration of the KG fields: Under the conformal symmetry of the modified KG equation, the curved spacetime metric, the former KG solution, and the fixed mass parameter are mapped to Minkowski spacetime, the corresponding KG solution, and the time-dependent mass term, respectively.}
    \label{fig:manifolds}
\end{figure*}

Eq.~\eqref{eq: FLRWtoFlat}  thus provides us with a transformation that relates a KG solution on an FLRW spacetime to the corresponding one on Minkowski spacetime (see Fig.~\ref{fig:manifolds} for a schematic illustration). 
Within the theory of a massive scalar field on FLRW spacetime, these two descriptions are indistinguishable. Note, however, that this is a consequence of restricting to this subtheory; if we were to introduce additional probes allowing us to access properties of the metric directly, it would of course be possible to differentiate between a situation with a curved and flat spacetime background. For example, measuring a non-vanishing Ricci scalar would provide evidence for the former case. In absence of such additional probes, however, we can effectively treat these two situations as physically equivalent. 

Overall, the set of transformations \eqref{eq: summary transformations conf} and the concrete example \eqref{eq: transformations FLRW} constitute symmetry operations 
that leave the action and the equation of motion, namely the KG equation in curved spacetime, invariant. This is an important feature for two reasons: Firstly, it is an instance of a symmetry transformation that relates different spacetimes rather than a transformation on a single spacetime. This paves the way to understanding superpositions of spacetimes in terms of symmetries. Secondly, representation theory is the natural language to understand symmetries in quantum theory. Having found a group representation thus gives us the necessary tools to tackle the problem on the quantum level. In particular, it will allow us to construct an explicit QRF transformation operator which maps between different, non-diffeomorphically related spacetimes.

\section{Superpositions of Spacetimes and Quantum Conformal Transformations} \label{sec: quantum case}

Let us now turn to the main objective of this work: elevating the classical symmetries studied above to the quantum level. First, let us note that the conformal invariance not only holds for the classical KG field but also for the quantized one. For a revision of the quantization of a scalar KG field, see App.~\ref{app: Quantization of the KG field}. Second, and most importantly, we now consider 
\emph{superpositions} of semiclassical states peaked around different conformally equivalent metrics while neglecting the quantum indefiniteness of geometry in each branch. Formally, we denote such a state by
\begin{align}
    \sum_i \alpha_i \ket{g_i},
\end{align}
where $\sum_i |\alpha_i|^2=1$. Note that the different $\ket{g_i}$ can be completely characterized by a single function $\Omega_i(x)\in C^2(\mathbb{R}^{3+1})$ and one representative of the conformal equivalence class, such that the Hilbert space accommodating the conformally related metrics is simply $\mathcal{H}^{(g)} \cong L^2(C^2(\mathbb{R}^{3+1}),\mu)$ with $\mu=\mathcal{D}\phi$ the measure on the space of twice differentiable real functions on $\mathbb{R}^{3+1}$.\footnote{Some caution is needed with the definition of this Hilbert space. We require the following two properties to hold on this Hilbert space: 1) The states assigned to two different functions $\Omega_1$, $\Omega_2 \in C^2(\mathbb{R}^{3+1})$ are orthogonal, that is, $\langle \Omega_1|\Omega_2\rangle=0$. 2) The state assigned to any function $\Omega \in C^2(\mathbb{R}^{3+1})$ is normalized, that is, $\langle \Omega|\Omega\rangle=1$. We denote the inner product on the Hilbert space $L^2(C^2(\mathbb{R}^{3+1}),\mu)$ by $\langle\varphi|\psi\rangle=\int \mathcal{D}\phi \varphi[\phi]\psi[\phi]$ where $\mu=\mathcal{D}\phi$ is the measure on the space of twice differentiable real functions on $\mathbb{R}^{3+1}$. While it is non-trivial to make the measure on this space well-defined, there are fruitful attempts in the literature on quantum field theory. An example is the 
construction based on the axioms of Osterwalder and Schrader (see Refs.~\cite{Ranard_2015, glimm1987} for more details). The same caveats hold for the definition of $\mathcal{H}^{(m)} \cong L^2(C^2(\mathbb{R}^{3+1}),\mu)$.} As a consequence, macroscopically different solutions $g_i$ correspond to orthogonal quantum states. We would expect such a superposition of spacetimes to arise in the transition of general relativity to a full theory of quantum gravity. In fact, near-future table-top experiments might be able to create such states in the regime of weak gravitational fields by placing a massive object in superposition of distinct locations \cite{aspelmeyer2021}. Let us state this expectation explicitly as one of three following assumptions that we make for the remainder of this work \cite{zych_2019, giacomini2021einsteins}:
\begin{enumerate}
    \item Macroscopically distinguishable gravitational fields are assigned orthogonal quantum states (where macroscopically distinguishable refers to being distinguishable by the measurement of macroscopic observables).
    \item Quantum field theory on curved spacetime holds for each well-defined spacetime.
    \item The superposition principle holds for such gravitational fields.
\end{enumerate}
These assumptions can be seen as a conservative extension of quantum theory and general relativity to the regime of macroscopic superpositions of geometries in that they preserve both the linearity of quantum theory and the equations of motion of general relativity. We would like to stress at this point that, when considering a superposition of geometries, we are not committing to any one particular interpretation of quantum theory -- how one makes metaphysical sense of states in quantum superposition will differ across the various interpretations but the present formalism is interpretation-neutral. Since we are extending the known theories of general relativity and quantum field theory on curved spacetime to superpositions of gravitational fields, we expect our model to break down in the same regime as these theories do -- that is, close to the singularities of general relativity and realms of extremely high curvature \cite{Bojowald_2008}. While we do not claim that our model is appropriate in these regimes, there exist concrete proposals in the quantum gravity and quantum cosmology literature that extend to these extreme scales while approximating a classical geometry and thus our second assumption outside of these regimes \cite{Ashtekar_2009}. Just as in QFT on curved spacetime, we also restrict to scenarios in which the backreaction of the matter fields on the geometry can be neglected.

In  the following, we want to explore the consequences of the above assumptions for the behavior of the quantized KG field on a superposition of conformally related spacetimes. While we lack a well-founded intuition for the latter, we can use the physical equivalence established in the previous sections to translate this description to one that is more easily understood in the language of QFTCS. In particular, our extension of the quantum reference frame formalism and the extended symmetry that we establish below show that the situation described above corresponds to one in which there is a quantum field with different mass terms in a superposition on a single spacetime background.

These two equivalent situations can be associated with two different perspectives or, equivalently, two different \emph{abstract} quantum reference frames (QRF). Note that when referring to QRFs in this context, we go beyond the usual interpretation as a physical system or even quantum coordinates. Instead, we take as a necessary condition that a QRF induces a specific factorization of the total Hilbert space in which the state of the frame ought to factorize out \cite{delaHamette2020}. By further specifying the state of the frame, both the states of the remaining systems and the frame change operator are uniquely determined. One can thus view the situation of spacetimes in superposition as being described relative to the frame associated to the mass of the KG field, while the description in terms of quantum fields in a mass superposition on a fixed spacetime background can be understood as \enquote{relative to the metric}. We can formalize this by introducing the states $\ket{\Psi}^{(m)}$ and $\ket{\Psi}^{(g)}$ \enquote{relative} to the mass and the metric respectively:
\begin{align}
    &\ket{\Psi}^{(m)}= \left(\sum_i \alpha_i \ket{g_i} \otimes \ket{\cb{\Phi}^{(g_i,m^2)}} \right)\otimes \ket{m^2}  \nonumber\\ \Leftrightarrow \ 
 &\ket{\Psi}^{(g)}= \ket{g} \otimes \left( \sum_i \alpha_i \ket{\cb{\Phi}^{(g,m_i^2)}} \otimes \ket{m_i^2} \right),\label{eq: equivalence}
\end{align}
where $\{g_i\}_i$ and $g$ on the one hand, and $\{m_i\}_i$ and $m$ on the other, belong to the same conformal class, as introduced in Eq.~\eqref{eq: summary transformations conf}\footnote{Since we restrict the codomain of the transformation to non-negative mass terms and thus physically meaningful situations, not all superpositions of spacetimes that are related by a conformal transformation can be \enquote{transformed away} into a superposition of an effective mass that is non-negative everywhere.}. Let us briefly explain the meaning of the various factors of the tensor product. Just as $\ket{g}$ denotes a semiclassical state peaked around the metric $g$, $\ket{m^2}$ indicates a semiclassical state for a given mass term governing the behavior of the KG field. A superposition of such (effective) mass terms could be understood as arising from the coupling to an external potential that is, itself, sourced by a quantum system and can thus be found in a superposition state. An alternative view would be to see it as effectively describing a superposition of different internal energy states (cf.~\cite{Castro_Ruiz_2017, zych_2011, Pikovski_2013}). The field itself is described by the general field state $|\Phi^{(g,m^2)}\rangle$, which can be expressed as a linear combination of eigenstates $|\phi^{(g,m^2)}\rangle$ of the field operators $\hat{\phi}^{(g,m^2)}$ [see \cite{Hatfield}, Ch.~10]. That is, $\hat{\phi}^{(g,m^2)} |\phi^{(g,m^2)}\rangle =\phi^{(g,m^2)} |\phi^{(g,m^2)}\rangle$. The field state depends on both the spacetime and the mass as both enter the equations of motion. Note that, as a consequence, the quantum field is always in a superposition, independently of the perspective (unless $\alpha_0 = 1$, $\alpha_i=0$ for $i>0$). Moreover, each $|\Phi^{(g,m^2)}\rangle$ already denotes the state of a quantum KG field, which itself can be expressed as a superposition of various Fock states. We go beyond standard QFTCS by considering a superposition of different such field states, labeled by different $(g,m^2)$. Note that we use the Schrödinger picture here to denote the quantum states $\ket{\Psi}^{(m)}$ and $\ket{\Psi}^{(g)}$ since this allows to straightforwardly define the frame change operator in the spirit of standard QRF transformations \cite{Giacomini_2019, delaHamette2020} in what follows.

We can now utilize the fact that the transformation between the abovementioned physical situations forms a representation of the symmetry group $(C^2(\mathbb{R}^{3+1}),\cdot)$ to immediately define the QRF transformation operator $\hat{\mathcal{S}}^{(m\to g)}: \mathcal{H}^{(m)} \to \mathcal{H}^{(g)}$, which maps between states relative to fixed $m$ and states relative to fixed $g$.\footnote{The operator $\hat{\mathcal{S}}^{(m\to g)}$ maps between isomorphic Hilbert spaces since in both spaces the states are characterized by real functions on $\mathbb{R}^{3+ 1}$: $m^2(x)$ in the case of the former and $\Omega(x)$ in the case of the latter. Note that the action of the operator $\hat{\mathcal{S}}^{(m\to g)}$ is restricted to superpositions of states for which $\{ \ket{g_i} \}_i$ belong to the same conformal class. This ensures that the codomain of the operator is indeed $\mathcal{H}^{(g)}$. We thus have 
$\mathcal{H}^{(m)}\cong \mathcal{H}^{(g)} \cong L^2(C^2(\mathbb{R}^{3+1}),\mu)$. Together with $\mathcal{F}$ being bijective, this implies that $\hat{\mathcal{S}}^{(g\to m)}$ is invertible.} In particular, we require $\hat{\mathcal{S}}^{(m \to g)} \ket{\Psi}^{(m)} = \ket{\Psi}^{(g)}.$ The action of this operator on an arbitrary basis state in $\mathcal{H}^{(m)}$ is given by
\begin{align}
 &\hat{\mathcal{S}}^{(m \to g)} \left( \ket{g_i} \ket{\phi^{(g_i,m^2)}}\ket{m^2} \right) \nonumber \\ =  &\underset{\ket{g}}{\underbrace{\ket{\mathcal{F}_g(\Omega_i)[g_i]}}} \ket{\mathcal{F}_\phi(\Omega_i)[\phi^{(g_i,m^2)}]}\ket{\mathcal{F}_m(\Omega_i)[m^2]},\label{eq: QRFchangeOperator}
\end{align}
where $|\mathcal{F}_\phi\cb{(\Omega_i)}[\phi^{(g_i,m^2)}]\rangle=|\phi^{(\mathcal{F}_g(\Omega_i)[g_i],\mathcal{F}_m(\Omega_i)[m^2])}\rangle$.
In general, we can map to any fixed metric $g$ in the conformal equivalence class. The choice of a specific $g$ selects the reference frame and associated Hilbert space $\mathcal{H}^{(g)}$. As a consequence, this gives rise to a unique frame change operator and a unique state on the right-hand side of Eq.~\eqref{eq: QRFchangeOperator}.
Similarly, if we want to go from a state relative to a fixed spacetime background $g$ to one in which the field has a definite mass term $m$, we use the inverse operator $\hat{\mathcal{S}}^{(g \to m)} = (\hat{\mathcal{S}}^{(m \to g)})^\dagger$, which acts on basis states as
\begin{align}
    &\hat{\mathcal{S}}^{(g \to m)}\left(\ket{g} \otimes \ket{\phi^{(g,m_i^2)}}\ket{m_i^2} \right) \nonumber \\ = &\ket{\mathcal{F}_g(\Omega_i^{-1})[g]} \ket{\mathcal{F}_\phi\cb{(\Omega_i)}[\phi^{(g,m_i^2)}]}\underset{\ket{m}}{\underbrace{\ket{\mathcal{F}_m(\Omega_i^{-1})[m_i^2]}}},
\end{align}
where we made use of the representation property $\mathcal{F}^{-1}(\Omega) = \mathcal{F}(\Omega^{-1})$ \footnote{Note that one can only map to a situation with superposed spacetimes and \emph{one} fixed mass term $m(x)$ if, in the original frame, the mass terms $m_i^2(x)$ in superposition are all of a specific functional form, that is, they are the image of a single (spacetime dependent) mass term $m^2(x)$ under Eq.~\eqref{eq: summary transformations conf}.}.

\begin{figure*}
    \centering
    \subfigure[A KG field with fixed mass term living on a semiclassical superposition of different conformally flat spacetimes.]
    {
        \centering
        \hfill
        \includegraphics[scale=0.23]{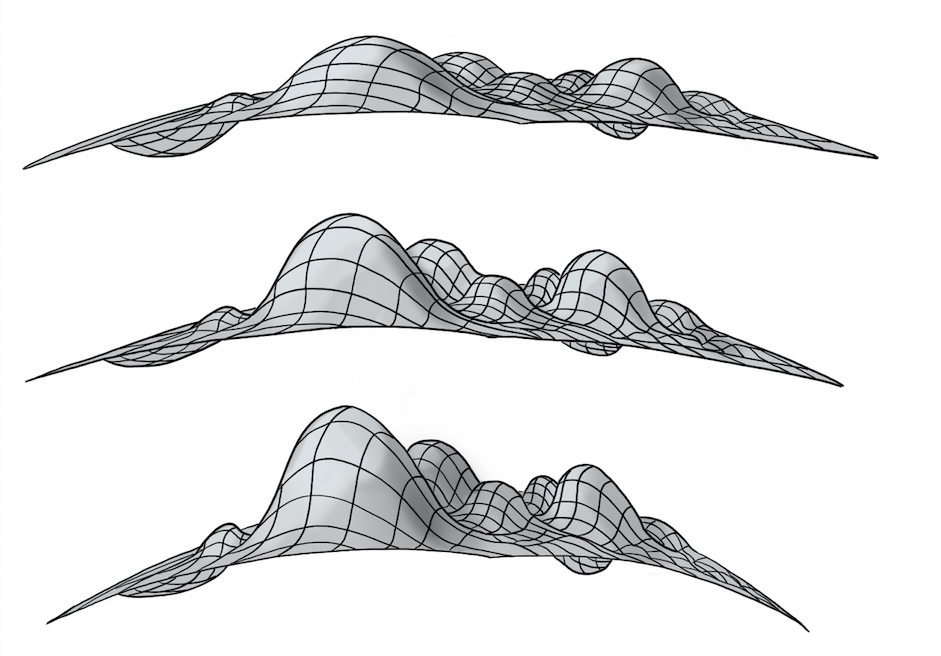}
        \label{fig: leftmanifoldsSP}
    }
    \qquad
    \subfigure[A KG field with spacetime-dependent mass term in superposition on a definite flat spacetime background.]
    {
        \centering
        \hfill
        \includegraphics[scale=0.23]{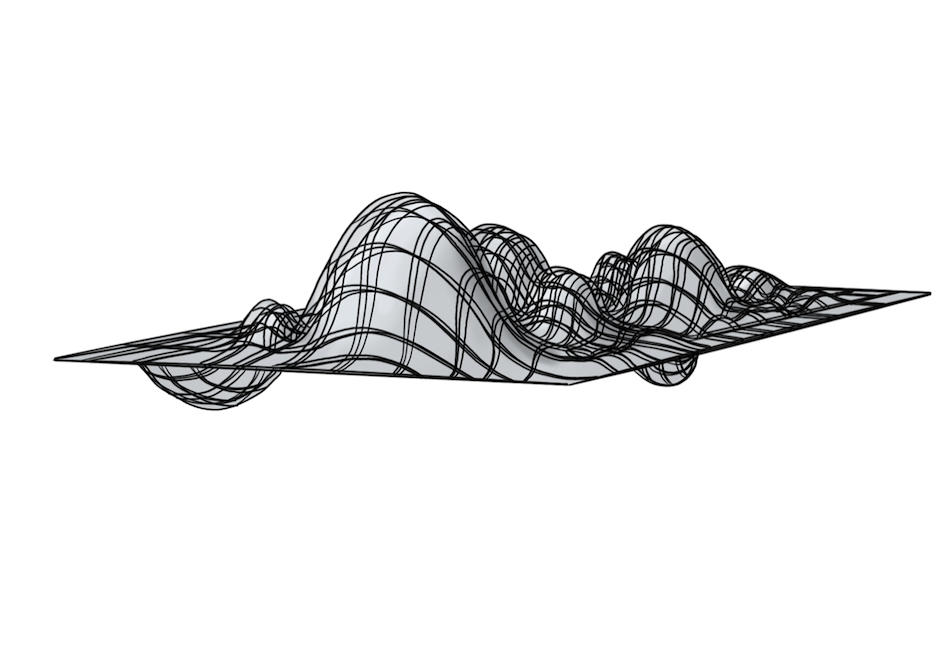}
         \label{fig: rightmanifoldsSP}
    }
    \caption{A schematic pictorial illustration of the two corresponding situations related by a quantum conformal transformation. The transformation operator $\hat{\mathcal{S}}^{(m\to g)}$, defined in Eq.~\eqref{eq: QRFchangeOperator}, maps the situation in (a) to the corresponding one in (b).}
    \label{fig:manifoldsSP}
\end{figure*}

\begin{figure*}
    \centering
    \includegraphics[scale=0.3]{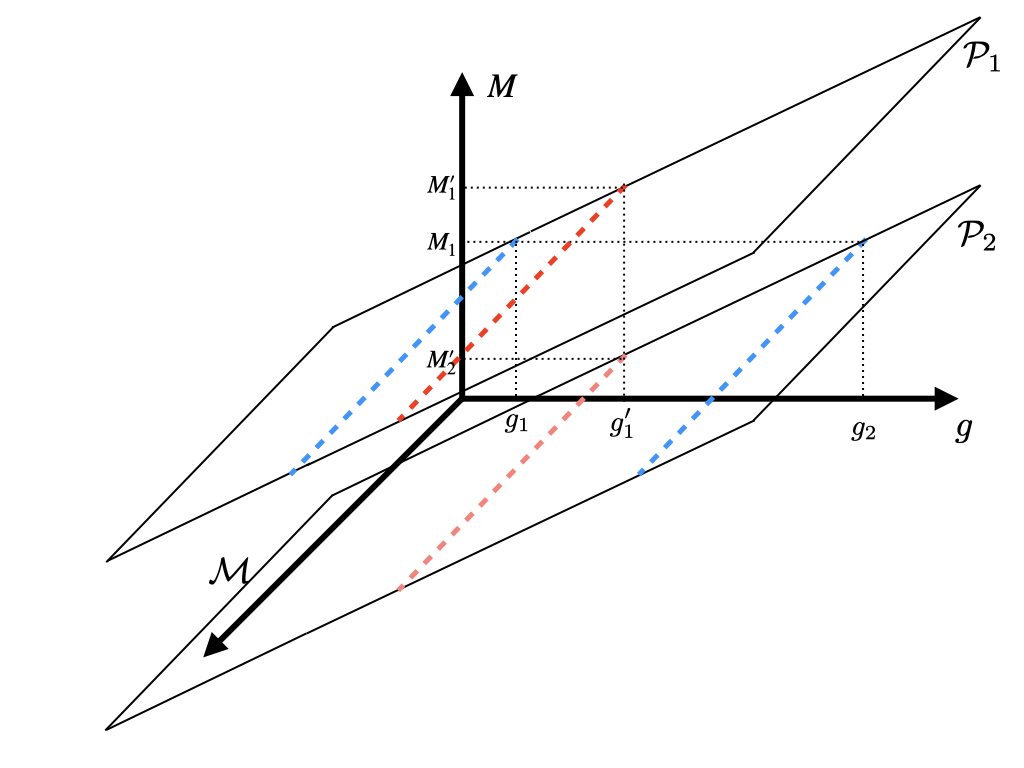}
    \caption{An abstract illustration of the different scenarios related by a \emph{quantum conformal transformation}. The axis labeled by $\mathcal{M}$ represents different points on a spacetime manifold. The axis labeled by $g$ represents different values of the metric. The axis labeled by $M$ represents different values of the modified mass term. The broken colored lines depict field configurations. All field configurations lying on the tilted plane $\mathcal{P}_1$ are related by an element of the symmetry group and can therefore be seen as equivalent. An analogous statement holds for the tilted plane $\mathcal{P}_2$. In the figure, we illustrate two equivalent situations. The dotted lines in blue illustrate a field configuration in a quantum superposition of two different, conformally related metrics, $g_1$ and $g_2$. Here, the value of the mass term is equal to $M_1$ for both amplitudes. By performing a quantum conformal transformation, we can \enquote{slide} the configuration  along each plane in a quantum-controlled manner. The end result is the situation represented by the orange dotted lines. In this new quantum field state, the mass term is in a quantum superposition of the values $M^\prime_1$ and  $M^\prime_2$. However, the metric is no longer in a superposition and has the definite value $g^\prime_1$ for the two amplitudes. Therefore, by performing a quantum conformal transformation, we can understand the physics of quantum fields in a superposition of semiclassical spacetimes in terms of the physics of quantum fields with a superposition of masses on a fixed spacetime.}
    \label{fig:symmetryplanes}
\end{figure*}

This operator defines a \emph{quantum conformal operator} and allows us to map between situations with conformally related spacetimes in superposition to scenarios in which we have quantum fields with different mass terms in superposition on a fixed background metric (see Fig.~\ref{fig:manifoldsSP} for a schematic picture and Fig.~\ref{fig:symmetryplanes} for an abstract illustration). Moreover, assuming that the conformal invariance of the KG equation with modified mass extends to superpositions of conformally equivalent spacetimes, it forms a symmetry transformation, which only alters the description while leaving the physical situation unchanged. The present work thus extends the quantum symmetry principle proposed in Ref.~\cite{delaHamette2021falling} -- there, situations with a gravitating source in a spatial superposition were related to ones in which the spacetime metric is definite while the probe particles are in superposition.

\section{Cosmological particle production} \label{sec: particle production}

We now use the tools constructed above to study the equivalent of cosmological particle production in Minkowski spacetime with an effective mass term in superposition, which will allow us to identify a quantum conformally invariant observable. Note that we change to the Heisenberg picture in this section since this significantly simplifies the description.

Cosmological particle production in curved spacetime is by now a well-studied phenomenon. In App.~\ref{app: cosmological particle production}, we briefly remind the reader of the main points in its derivation, following \cite{Ford_2021}. In short, in a spacetime with a time-dependent scale factor that has constant but different values $a_{in}$ and $a_{out}$ in the asymptotic past and future, the expected number of particles in mode $k$ at future infinity with respect to the vacuum at past infinity is 
\begin{align}
   \expval{\hat{N}_k^+}{0_-}_g = \sum_{k'}|\beta_{k'k}|^2.
\end{align}

\begin{figure*}
    \centering
    \subfigure[Time-dependence of the scale factor $a(\tau)$ in FLRW spacetime from initial value $a_{in}$ at $\tau\to-\infty$ until final value $a_{out}$ at $\tau\to+\infty$.]
    {
        \centering
        \hfill
        \includegraphics[scale=0.2]{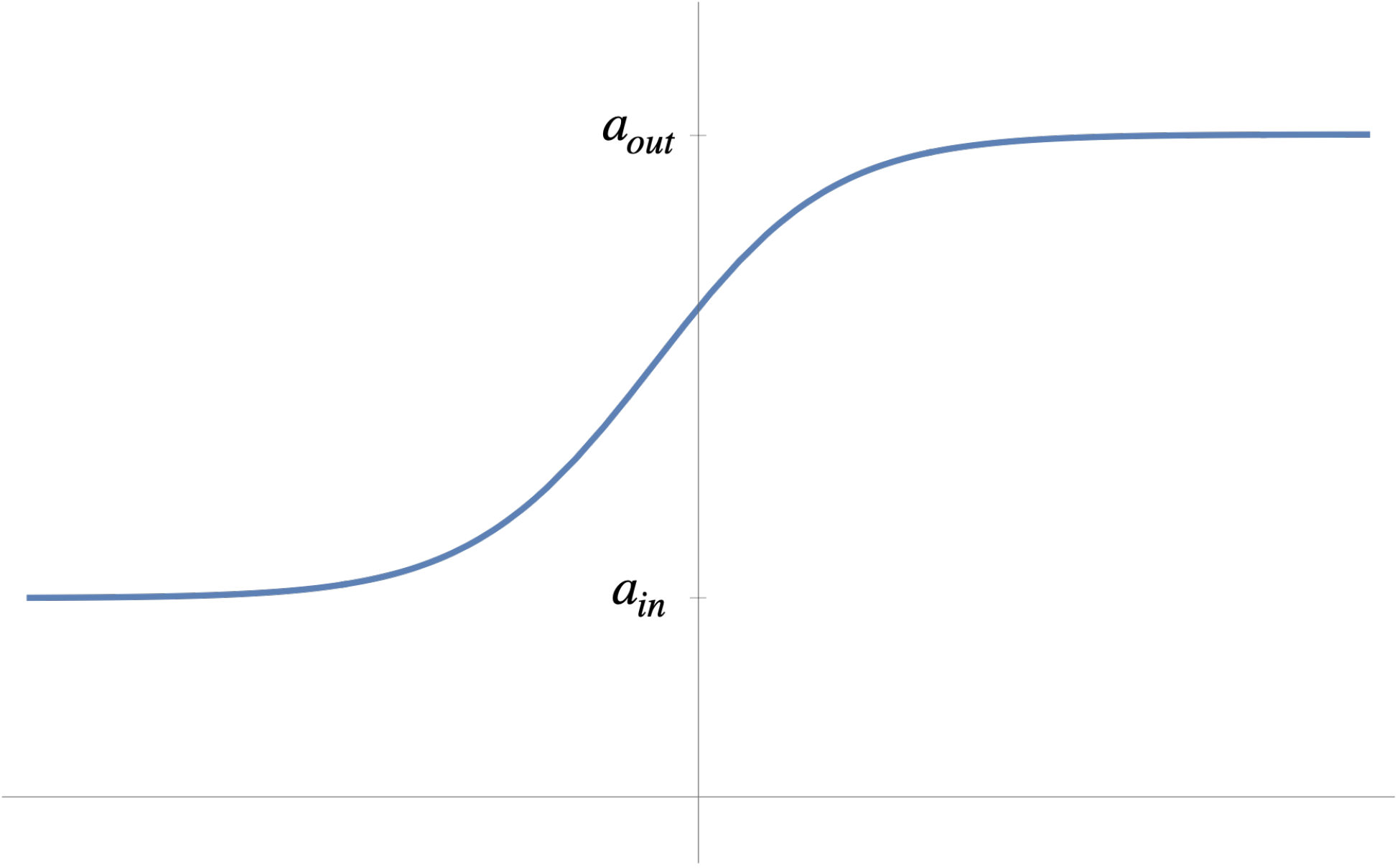}
        \label{fig: scalefactor}
    }
    \qquad 
    \subfigure[Time-dependence of the mass term $M^2(\tau)$ in modified Minkowski from initial value $a_{in}^2m^2$ at $\tau\to-\infty$ until final value $a_{out^2}m^2$ at $\tau\to+\infty$.]
    {
        \centering
        \hfill
        \includegraphics[scale=0.2]{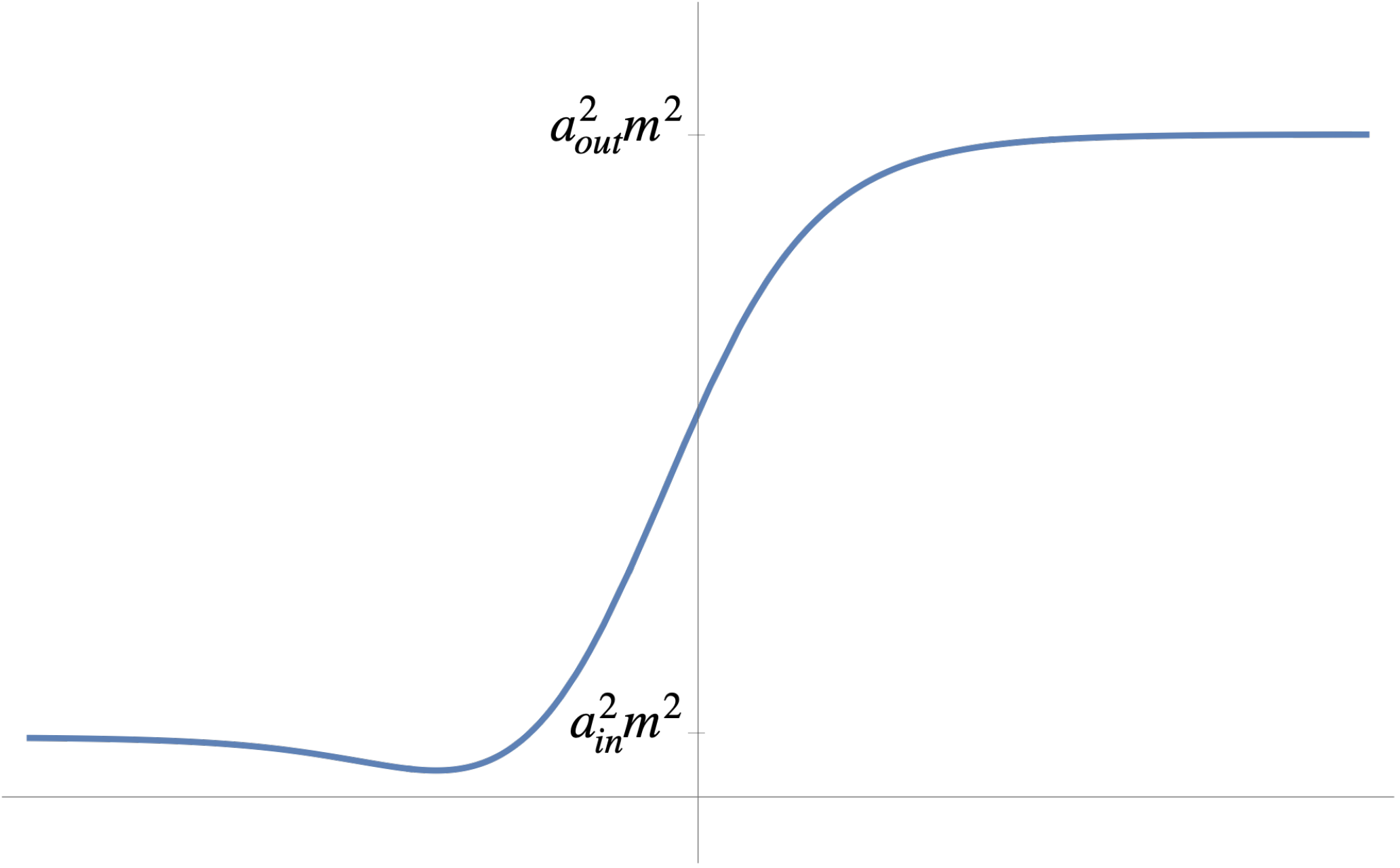}
         \label{fig: massterm}
    }
    \caption{The quantum conformal symmetry between FLRW and modified Minkowski spacetime manifests itself in the equivalence of the creation and annihilation operators and thus all quantities derived from them. As a consequence, we observe the phenomenon of particle production not only in curved spacetime but also in modified Minkowski. While the former is due to the time-dependence of the scale factor $a(\tau)$, the latter derives from a time-dependent mass term $M^2(\tau)$.}
    \label{fig:aM}
\end{figure*}

This holds equally in Minkowski spacetime with modified KG mass. In the asymptotic regions $\tau \to \pm \infty$, the creation and annihilation operators in modified Minkowski are the same as in the corresponding curved spacetime. To see this, consider two general solutions $\tilde{\phi} = a\phi$ and $\tilde{\varphi} = a\varphi$ of the Minkowski KG equation, related to the curved solutions via the scale factor $a$, through transformation \eqref{eq: summary transformations conf}. Using the definitions \eqref{eq: innerProd} and \eqref{eq: innerProdMink} of the pseudo inner products, one can show that
\begin{align}
    (\tilde{\phi},\tilde{\varphi})_\eta = (a\phi, a\varphi)_\eta = (\phi,\varphi)_g.
\end{align}
Thus, it follows that classically, the coefficients of the field expansion are related by 
\begin{align}
    \tilde{a}_k = (\tilde{u}_k,\tilde{\phi})_\eta = (u_k,\phi)_g = a_k,\\
    \tilde{a}_k^\ast = (-\tilde{u}^\ast_k,\tilde{\phi})_\eta = (-u_k^\ast,\phi)_g = a_k^\ast.
\end{align}
Analogously, one can show that $\tilde{b}_k = b_k$ and $\tilde{b}^\ast_k=b^\ast_k$. 
As a consequence, upon quantization, also the creation and annihilation operators, the number operators and the vacuum states agree:
\begin{align}
    \hat{a}_k^\dagger =\hat{\tilde{a}}_k^\dagger &\text{, }\hat{a}_k =\hat{\tilde{a}}_k\\
    \hat{b}_k^\dagger=\hat{\tilde{b}}_k^\dagger&\text{, } \hat{b}_k=\hat{\tilde{b}}_k,\\ \hat{N}_k^{+,-}&=\hat{\tilde{N}}_k^{+,-},\\  \ket{0_{+,-}}_g &= \ket{0_{+,-}}_\eta.
\end{align}
This shows that the number operator $\hat{N}_k^{+,-}$ is invariant under conformal transformations as described in Eq.~\eqref{eq: FLRWtoFlat}. Furthermore, since the Bogoliubov transformations between $\hat{\tilde{a}}_k$ and $\hat{\tilde{b}}_k$ as well as $\hat{\tilde{a}}^\dagger_k$ and $\hat{\tilde{b}}^\dagger_k$  in modified Minkowski are also the same, it follows that 
\begin{align}
    \expval{\hat{N}_k^+}{0_-}_g &= \sum_{k'}|\beta_{k'k}|^2 \nonumber \\ &= \sum_{k'}|\tilde{\beta}_{k'k}|^2 = \expval{\hat{\tilde{N}}_k^+}{0_-}_\eta.
\end{align}
Thus, the same mean number of particles is produced in modified Minkowski spacetime equipped with a time-dependent mass $M^2(\tau) = a^2(\tau)m^2 - \frac{\ddot{a}(\tau)}{a(\tau)}$. While the particle production on FLRW is due to the spacetime curvature between the asymptotic regions, the effect in Minkowski spacetime emerges because of the time-dependence of the mass term, which can also be understood as an additional potential for the KG field (see Fig.~\ref{fig:aM}).

Let us now show that this effect arises equally
in the case of a superposition of different FLRW metrics with a different numbers of particles produced in each branch. Firstly, it follows immediately from the invariance of the number operator under conformal transformations that it is also invariant under \emph{quantum conformal transformations} since these amount to applying an independent conformal transformation in each branch. Secondly, to see that the mean number of particles in each branch remains the same under these transformations, let us consider an arbitrary superposition of conformally related metric states $\sum_i\alpha_i \ket{g_i}$. In each branch $i$ of the superposition, we can apply the same argument as above to obtain the following correspondence:
\begin{align}
    \expval{\hat{N}_k^+}{0_-}_{g_i} = \expval{\hat{\tilde{N}}_{k,i}^+}{0_-}_{\eta_i},
\end{align}
where $\hat{\tilde{N}}_{k,i}^+$ and $\ket{0_-}_{\eta_i}$ denote the number operator and vacuum state, respectively, in modified Minkowski spacetime with effective KG mass $m_i$. By linearity, the expectation value of a superposition of particle numbers then remains the same under quantum conformal transformations. That is, the phenomenon can arise from either a superposition of geometries or an effective mass in superposition. This shows that the equivalence between curved spacetime and modified Minkowski sheds light in both directions -- not only does the map to flat spacetime with an effective mass improve our understanding of a superposition of geometries; we can also use results from QFTCS to infer analogous effects in flat spacetime for a KG field with an effective mass in superposition. Generally, for a given problem, one can pick the frame more suited for one's particular question.

\section{Conclusion} \label{sec: conclusion}

We started this work by asking the question of how fields and particles would behave in a universe that is expanding in a quantum superposition of different Hubble rates. By exploiting a particular symmetry of the Klein-Gordon equation, we showed that this situation corresponds to one in which a field in superposition of different spacetime-dependent effective masses inhabits a definite Minkowski spacetime. More concretely, we constructed an operator $\hat{\mathcal{S}}$, which maps between quantum states describing these two situations. This allowed us to formalize a new extended symmetry principle: an invariance under quantum conformal transformations, which manifests in the equivalence of states that are related by the action of $\mathcal{\hat{S}}$. Besides providing a deeper intuition for the meaning of quantum fields on geometries in superposition, we further utilized this symmetry to derive the equivalent of cosmological particle production in Minkowski spacetime with modified mass in superposition. We considered an FLRW spacetime with a general time-dependent scale factor $a(\tau)$, which approaches constant values in the far past and future. By relating the creation and annihilation operators between curved and flat spacetime in the asymptotic limit, we found that the number operator $\hat{N}^{+,-}_k$ represents a conformally invariant observable. Consequently, the phenomenon of particle production in curved spacetime is reproduced in  Minkowski with modified KG mass term. This result straightforwardly extends to superpositions of conformally related spacetimes or their Minkowski equivalent, with different numbers of particles produced in each branch. This being the first concrete application of the principle of \emph{quantum conformal invariance}, we are positive that it provides a useful tool to uncover new phenomena at the interface of quantum physics and gravity.

The present work extends the results of Ref.~\cite{delaHamette2021falling} in several aspects. First, we have now gone beyond quantum coordinate transformations by mapping between any conformally related and thus in general diffeomorphically inequivalent metrics. The strength of these transformations is that they allow one to map a superposition of the latter to a definite background metric.
Second, to our knowledge, this is the first instance of treating quantized scalar fields on superposed spacetimes in the context of the quantum reference frame program. We are thus probing superpositions of geometries with quantum fields. On a different note, we go beyond the conformal invariance of the non-minimally coupled KG equation, which has been extensively discussed in the literature [cf.~e.g.~\cite{Falciano_2012, Hammad_2021}]. Firstly, we offer a new perspective by viewing the description in terms of fields on a modified Minkowski spacetime as a physically instantiated situation instead of a mere computational tool \cite{Pereira_2010,ParticlePhysicsCosmology}. In doing so, we take seriously the spacetime-dependence of the effective mass term. Alternatively, one can interpret the evolving KG mass parameter on flat spacetime as an additional potential for the field \cite{Agullo_2013}. Secondly, utilizing the group properties of the conformal transformations, we are able to promote the latter to \emph{quantum} conformal transformations, which act on superpositions of geometries.

Let us further point out that in identifying the presently discussed symmetry principle, it was crucial to (a) restrict to a subtheory of quantum field theory, namely that of a massive KG field, and (b) to allow for transformations of quantities that are usually left untouched, specifically the mass term. Often, and in particular in the context of quantum gravity research, the focus is mainly placed on the symmetries of the most general theories, that is, the diffeomorphism invariance of general relativity and the gauge symmetries of quantum field theory. While this is important if one wants to identify the symmetries of a full theory of quantum gravity, more immediate progress may be made by restricting the range of application but enlarging the set of allowed transformations leaving the equations of motion invariant. This approach has the potential to uncover a wide range of new or previously disregarded symmetries. By extending these to the quantum level, we can enhance our understanding of specific scenarios at the intersection of quantum physics and gravity without requiring a full theory of quantum gravity that captures them. 

Further directions for future work include going beyond superpositions of semiclassical metrics and taking into account the quantum indefiniteness of the metrics. This involves a number of technical steps which we were able to circumvent here, such as the Hilbert space structure of more general metrics. Moreover, we are hopeful that the aforementioned strategies utilizing symmetry principles can be used to extend the formalism of abstract quantum reference frame transformations to more general superpositions of metrics, not necessarily belonging to the same conformal class. Thinking further ahead, one might want to treat spacetime metrics equipped with the quantum fuzziness expected in
quantum gravity. This is likely to require the construction of a framework of QRFs for fields. If successful, the latter would allow to change into a quantum frame in which the metric field still becomes definite, shifting any genuine quantum fluctuations into degrees of freedom of the probe systems, that is, the quantum fields inhabiting the spacetime.

Finally, it would be interesting to see to what extent this effective correspondence can be investigated experimentally. Analogue experiments \cite{Visser_2005, Barcelo_2005} seem to be the ideal playground on which to test both the phenomenon of particle production in flat and superposed curved spacetimes and the quantum conformal symmetry principle. However, in light of recent philosophical discussions \cite{Crowther_2018}, one should be careful not to confound empirical results of analogue experiments with the verification of hypotheses in quantum gravity setups. Performing the experiment with a KG field on Minkowski with mass terms in superposition, however, would bear insights by itself, since it can be seen as an actual \emph{realization} of the phenomenon of particle production in flat spacetime rather than an analogue experiment. While actually testing the symmetry principle in the context of FLRW spacetimes in superposition seems experimentally out of reach, the strength of the present work lies in its potential to enhance our intuition, sharpen our technical tools, and identify guiding principles to deal with situations at the interface between quantum theory and gravity.

\begin{acknowledgments}
We thank Ali Akil, Marios Christodoulou, Luis Cortés-Barbado, Duc Viet Hoang, Thomas Mieling, and Wolfgang Wieland for helpful discussions. E.C.-R. is supported by an ETH Zurich Postdoctoral Fellowship and acknowledges financial support from the Swiss National Science Foundation (SNSF) via the National Centers of Competence in Research QSIT and SwissMAP, as well as the project No. 200021\_188541. We acknowledge financial support by the Austrian Science Fund (FWF) through BeyondC (F7103-N48), the Austrian Academy of Sciences (ÖAW) through the project ``Quantum Reference Frames for Quantum Fields" (ref.~IF 2019\ 59\ QRFQF), the European Commission via Testing the Large-Scale Limit of Quantum Mechanics (TEQ) (No. 766900) project, the Foundational Questions Institute (FQXi) and the Austrian-Serbian bilateral scientific cooperation no.~451-03-02141/2017-09/02. This publication was made possible through the support of the ID 61466 grant from the John Templeton Foundation, as part of The Quantum Information Structure of Spacetime (QISS) Project (qiss.fr). The opinions expressed in this publication are those of the authors and do not necessarily reflect the views of the John Templeton Foundation.
\end{acknowledgments}

\nocite{apsrev42Control}
\bibliography{bibliography}
\bibliographystyle{apsrev4-2.bst}

\onecolumngrid

\appendix\label{sec: Supplementary Note}

\section{Conformal Invariance of the KG Equation and Representation Properties} \label{app: invariance conformal}
Consider two metrics $(g_1)_{ab}=\Omega^2(g_2)_{ab}$ belonging to the same conformal class. Starting from the KG equation in spacetime $g_1$
\begin{align}
    \square_{g_1}\phi_1(x)=m^2_1\phi_1(x),\label{eq: KG eqt g1}
\end{align}
where $x$ denotes a four-vector, we want to find a field $\phi_2(x)$ and mass term $m^2_2(x)$ such that 
\begin{align}
    \square_{g_2}\phi_2(x)=m^2_2(x)\phi_2(x).\label{eq: KG eqt g2 app}
\end{align} 
In order to do so, we follow Wald [\cite{Wald_gr}, pp. 445-446] in expressing the covariant derivative with respect to $g_2$ in terms of that of $g_1$:
\begin{align}
    \nabla^{(2)}_a\omega_b = \nabla^{(1)}_a\omega_b-C^c_{ab}\omega_c = \nabla^{(1)}_a\omega_b-(2\delta^c_{(a}\nabla^{(1)}_{b)}(\ln{\Omega})-(g_1)_{ab}(g_1)^{cd}\nabla^{(1)}_d(\ln{\Omega}))\omega_c.
\end{align}
By replacing abstract with spacetime indices, it follows that 
\begin{align}
    \square_{g_2} \phi_2(x)&=(g_2)^{\mu\nu} \nabla^{(2)}_\mu\nabla^{(2)}_\nu \phi_2(x) = \frac{1}{\Omega^2}(g_1)^{\mu\nu}(\nabla^{(1)}_\mu \partial_\nu \phi_2-C^\rho_{\mu\nu}\partial_\rho \phi_2) \nonumber\\
    &=\frac{1}{\Omega^2}(g_1)^{\mu\nu} \left[\nabla^{(1)}_\mu\nabla^{(1)}_\nu \phi_2-(2\delta^\rho_{(\mu}\nabla^{(1)}_{\nu)}(\ln{\Omega})-(g_1)_{\mu\nu}(g_1)^{\rho\sigma}\nabla^{(1)}_\sigma(\ln{\Omega})) \partial_\rho \phi_2 \right] \nonumber\\
    &=\frac{1}{\Omega^2} \left[ \square_{g_1} \phi_2 - (2(g_1)^{\rho\nu}\nabla^{(1)}_{\nu}(\ln{\Omega})+4(g_1)^{\rho\sigma}\nabla^{(1)}_\sigma(\ln{\Omega})) \partial_\rho \phi_2 \right] \nonumber\\
    &=\frac{1}{\Omega^2} \left[ \square_{g_1} \phi_2 + 2(g_1)^{\mu\nu}\nabla^{(1)}_{\mu}(\ln{\Omega}) \partial_\nu \phi_2\right].
\end{align}
If we now make the ansatz $\phi_2=\Omega^{-1} \phi_1$, then
\begin{align}
    &\square_{g_2} \phi_2(x) = \frac{1}{\Omega^2}\left[ (g_1)^{\mu\nu} \nabla^{(1)}_\mu\partial_\nu (\Omega^{-1}\phi_1)+2(g_1)^{\mu\nu}\nabla^{(1)}_\mu (\ln{\Omega})\partial_\nu(\Omega^{-1}\phi_1)\right] \nonumber\\
    &= \frac{1}{\Omega^2}(g_1)^{\mu\nu}\big[ \partial_\mu\left( (\partial_\nu \Omega^{-1}) \phi_1 +\Omega^{-1}\partial_\nu\phi_1\right)\vk{-}(\Gamma^{(1)})^\rho_{\mu\nu}\left( (\partial_\rho \Omega^{-1})\phi_1+\Omega^{-1}(\partial_\rho\phi_1) \right) \nonumber\\ 
    & \hspace{1cm}+2\partial_\mu (\ln{\Omega})\left( (\partial_\nu\Omega^{-1})\phi_1+\Omega^{-1}(\partial_\nu\phi_1)\right)\big] \nonumber\\
    &= \frac{1}{\Omega^3}(g_1)^{\mu\nu}\left(\partial_\mu\partial_\nu \phi_1\vk{-}(\Gamma^{(1)})^\rho_{\mu\nu}\phi_1\right)+\frac{1}{\Omega^2}(g_1)^{\mu\nu}\left[\partial_\mu\partial_\nu(\Omega^{-1})+2\frac{\partial_\mu\Omega}{\Omega}(\partial_\nu\Omega^{-1})\vk{-}(\Gamma^{(1)})^\rho_{\mu\nu}(\partial_\rho\Omega^{-1})\right] \phi_1 \nonumber\\
    &=\frac{1}{\Omega^3} \left[ \square_{g_1}\phi_1-(g_1)^{\mu\nu}\left( \frac{\partial_\mu\partial_\nu\Omega}{\Omega}\vk{-} (\Gamma^{(1)})^\rho_{\mu\nu}\frac{\partial_\rho\Omega}{\Omega} \right) \right]\phi_1.
\end{align}
Using Eq.~\eqref{eq: KG eqt g1}, this becomes
\begin{align}
    \square_{g_2}\phi_2(x) &= \frac{1}{\Omega^3} \left[ m_1^2\phi_1-(g_1)^{\mu\nu}\left( \frac{\partial_\mu\partial_\nu\Omega}{\Omega}\vk{-} (\Gamma^{(1)})^\rho_{\mu\nu}\frac{\partial_\rho\Omega}{\Omega} \right) \right] \phi_1 \nonumber \\
    &= \frac{1}{\Omega^2} \left[ m_1^2\phi_1-(g_1)^{\mu\nu}\left( \frac{\partial_\mu\partial_\nu\Omega}{\Omega}\vk{-} (\Gamma^{(1)})^\rho_{\mu\nu}\frac{\partial_\rho\Omega}{\Omega} \right) \right]\phi_2,
\end{align}
which satisfies Eq.~\eqref{eq: KG eqt g2 app}, given that we define
\begin{align}
    m_2^2(x) \equiv \frac{1}{\Omega^2(x)} \left[ m_1^2\phi_1-(g_1)^{\mu\nu}(x)\left( \frac{\partial_\mu\partial_\nu\Omega(x)}{\Omega(x)}\vk{-} (\Gamma^{(1)})^\rho_{\mu\nu}\frac{\partial_\rho\Omega(x)}{\Omega(x)} \right) \right].
\end{align}

The transformations of $g_1$, $\phi_1$ and $m_1^2$ are summarized in Eq.~\eqref{eq: summary transformations conf} and repeated here for ease of reading:
\begin{align}
    \begin{pmatrix}
     (g_1)_{ab} \\ \phi_1(x) \\ m_{1}^2
    \end{pmatrix}
    \to \left[  \begin{pmatrix}
     \mathcal{F}_g(\Omega)[(g_1)_{ab}] \\ \mathcal{F}_\phi(\Omega)[\phi_1(x)] \\ \mathcal{F}_m(\Omega)[m_{1}^2]
    \end{pmatrix} \right]=
    \begin{pmatrix}
     \Omega^2(x) (g_1)_{ab} \\ \Omega^{-1}(x)\phi_1(x) \\ \frac{1}{\Omega^2(x)}\left[ m_1^2-(g_1)^{\mu\nu}(x) \left( \frac{\partial_\mu \partial_\nu \Omega(x)}{\Omega(x)} \vk{-} (\Gamma^{(1)})_{\mu\nu}^\rho \frac{\partial_\rho \Omega(x)}{\Omega(x)} \right) \right]
    \end{pmatrix} 
    = \begin{pmatrix}
     (g_2)_{ab} \\ \phi_2(x) \\ m_{2}^2(x)
    \end{pmatrix}.
\end{align}
While it is completely straightforward to see that that the transformations $\mathcal{F}_g$ of $g_1$ and $\mathcal{F}_\phi$ of $\phi_1$ form a representation of the group $(C^2(\mathbb{R}^{3+1}),\cdot)$, let us show this explicitly for the mass term $m_1^2$. It is clear that $\mathcal{F}_m(\Omega=1)=\mathrm{Id}$ and $\mathcal{F}_m(\Omega^{-1})=\mathcal{F}_m(\Omega)^{-1}$. For the composition rule, consider the metrics $(g_2)_{\mu\nu}=(\Omega_{12})^2(g_1)_{\mu\nu}$ and $(g_3)_{\mu\nu}=(\Omega_{23})^2(g_2)_{\mu\nu}$. Note that the Christoffel symbols of conformally related metrics are related by 
\begin{align}
    \tilde{\Gamma}^k_{ij}=\Gamma^k_{ij}+\left(\delta^k_i \frac{\partial_j\Omega}{\Omega}+\delta^k_j \frac{\partial_i\Omega}{\Omega}-g_{ij}g^{kl}\frac{\partial_l\Omega}{\Omega}\right).
\end{align}
Now, it follows that
\begin{align}
    &\F_m[\Omega_{23}] \F_m[\Omega_{12}]m_1^2 = \F_m[\Omega_{23}] \frac{1}{(\Omega_{12})^2}\left[ m_1^2-(g_1)^{\mu\nu} \left( \frac{\partial_\mu \partial_\nu \Omega_{12}}{\Omega_{12}} \vk{-} (\Gamma^{(1)})_{\mu\nu}^\rho \frac{\partial_\rho \Omega_{12}}{\Omega_{12}} \right) \right] \nonumber\\
    &= \frac{1}{(\Omega_{23})^2}\left[ \frac{1}{(\Omega_{12})^2}\left( m_1^2-(g_1)^{\mu\nu} \left( \frac{\partial_\mu \partial_\nu \Omega_{12}}{\Omega_{12}} \vk{-} (\Gamma^{(1)})_{\mu\nu}^\rho \frac{\partial_\rho \Omega_{12}}{\Omega_{12}} \right) \right) -(g_2)^{\mu\nu} \left( \frac{\partial_\mu \partial_\nu \Omega_{23}}{\Omega_{23}} \vk{-} (\Gamma^{(2)})_{\mu\nu}^\rho \frac{\partial_\rho \Omega_{23}}{\Omega_{23}} \right) \right] \nonumber\\
    &=\frac{1}{(\Omega_{12}\cdot\Omega_{23})^2} \left[m_1^2-(g_1)^{\mu\nu} \left( \frac{\partial_\mu \partial_\nu \Omega_{12}}{\Omega_{12}} + \frac{\partial_\mu \partial_\nu \Omega_{23}}{\Omega_{23}}\vk{-} (\Gamma^{(1)})_{\mu\nu}^\rho \left(\frac{\partial_\rho \Omega_{12}}{\Omega_{12}}+\frac{\partial_\rho \Omega_{23}}{\Omega_{23}}\right) \right) \right. \nonumber\\ 
    & \hspace{0.5cm} \left. \left. \vk{+}(g_1)^{\mu\nu} \left( \delta_\mu^\rho \frac{\partial_\nu\Omega_{12}}{\Omega_{12}}+\delta_\nu^\rho \frac{\partial_\mu\Omega_{12}}{\Omega_{12}} \right. -(g_1)_{\mu\nu}(g_1)^{\rho\sigma}\frac{\partial_\sigma\Omega_{12}}{\Omega_{12}} \right)\frac{\partial_\rho\Omega_{23}}{\Omega_{23}} \right]\nonumber \\
    &= \frac{1}{(\Omega_{12}\cdot\Omega_{23})^2} \left[m_1^2- \left( \frac{(\partial^\mu \partial_\mu \Omega_{12})\Omega_{23}}{\Omega_{12}\Omega_{23}}+\frac{\Omega_{12}\partial^\mu \partial_\mu\Omega_{23}}{\Omega_{12}\Omega_{23}}+2\frac{\partial^\mu \Omega_{12}\partial_\mu \Omega_{23}}{\Omega_{12}\Omega_{23}} \right) \right. \nonumber\\
    & \hspace{0.5cm} \left. \vk{+}(g_1)^{\mu\nu} (\Gamma^{(1)})_{\mu\nu}^\rho \left( \frac{(\partial_\rho\Omega_{12})\Omega_{23}}{\Omega_{12}\Omega_{23}}+ \frac{\Omega_{12}\partial_\rho\Omega_{23}}{\Omega_{12}\Omega_{23}}\right)\right] \nonumber \\
    &= \frac{1}{(\Omega_{12}\cdot\Omega_{23})^2} \left[m_1^2-(g_1)^{\mu\nu} \left( \frac{\partial_\mu \partial_\nu(\Omega_{12}\Omega_{23})}{\Omega_{12}\Omega_{23}}\vk{-}(\Gamma^{(1)})_{\mu\nu}^\rho \frac{\partial_\rho(\Omega_{12}\Omega_{23})}{\Omega_{12}\Omega_{23}} \right) \right] \nonumber\\
    &= \F_m[\Omega_{12}\cdot\Omega_{23}]m_1^2.
\end{align}
Hence, we find that $\F_m[\Omega_{23}] \F_m[\Omega_{12}]=\F_m[\Omega_{12}\cdot\Omega_{23}]$ and thus $\F_m$ forms a representation of $(C^2(\mathbb{R}^{3+1}),\cdot)$. It is straightforward to conclude that the maps $\F_g$, $\F_\phi$ and $\F_m$ as given in Eq.~\eqref{eq: transformations FLRW} form representations of $(\mathbb{R},+)$ as well, since Eq.~\eqref{eq: transformations FLRW} can be seen as a special case of Eq.~\eqref{eq: summary transformations conf}.

\section{Conformal Invariance of KG Equation for FLRW} \label{app: invariance FLRW}
The line element of a spatially flat FLRW metric $g_{ab}$ is given by
\begin{align}
    ds^2&=-dt^2+a^2(t)\delta_{ij}dx^idx^j \nonumber\\
    &=a^2(t)(-d\tau^2+\delta_{ij}dx^idx^j),
\end{align}
where we take $a(t)=e^{Ht}$ and introduced the conformal time coordinate $\tau$ with $d\tau=a^{-1}(t)dt=e^{-Ht}dt$.
Consider a classical field $\phi$ that is a solution to the KG equation on $g_{ab}$ with mass term $m$:
\begin{align}
    \square_g \phi(t,x) = \left[ -(\partial_t +3H)\partial_t + e^{-2Ht}\Delta \right] \phi(t,x) = m^2 \phi(t,x).\label{eq: FLRW KG eq}
\end{align}
Now, let us show that $\tilde{\phi}(t,x) = e^{Ht} \phi(t,x)$ is a solution to the KG equation in flat spacetime $\eta_{ab}$ with modified mass. We have to be careful about the explicit form of $\square_\eta$ in $(t,x)$ coordinates though, since
\begin{align}
    \square_\eta = -\partial_\tau^2+\Delta = -e^{2Ht}(H\partial_t+\partial_t^2)+\Delta.
\end{align}
Thus, we have
\begin{align}
    \square_\eta \tilde{\phi}(t,x) &= \left[ -e^{2Ht} (H\partial_t + \partial_t^2)+\Delta \right] e^{Ht} \phi(t,x) \nonumber \\
    &= \left[-e^{3Ht} (H^2+H\partial_t +H^2+2H\partial_t+\partial_t^2)+e^{Ht}\Delta\right] \phi(t,x)\nonumber\\
    &= e^{3Ht}\left[ -(\partial_t +3H)\partial_t + e^{-2Ht}\Delta \right]\phi(t,x) -e^{3Ht}2H^2 \phi(t,x) \nonumber\\
    &= e^{3Ht}m^2 \phi(t,x)-e^{3Ht}2H^2 \phi(t,x)\nonumber\\
    &=e^{2Ht}(m^2-2H^2)\tilde{\phi}(t,x) \nonumber\\
    &\equiv M^2(t,x) \tilde{\phi}(t,x),
\end{align}
where, going from the third to the fourth line, we used Eq.~\eqref{eq: FLRW KG eq}. In the last line, we find the spacetime-dependent mass term $M^2(t,x) = e^{2Ht}(m^2-2H^2)$.

\section{Quantization of the KG field}\label{app: Quantization of the KG field}
We now proceed to quantize the KG field, once for curved spacetime, followed by a quantization in modified Minkowski. Following the standard procedure of Fock quantization \cite{Ford_2021} will allow us to define an explicit frame change operator in Sec.~\ref{sec: quantum case} and to consider the concrete phenomenon of particle production in Sec.~\ref{sec: particle production}. In curved spacetime, we first need to perform a $3+1$ split of the spacetime into $M = \Sigma \cross \mathbb{R}$ in order to define a conjugate pair of field and momentum. We denote by $h_{ab}$ the induced metric on the spatial hypersurface $\Sigma$ and by $n^a$ the corresponding future directed unit normal vector. Now, define the pseudo inner product (Hermitian form)
\begin{align}
    (\phi_1,\phi_2)_g=i\int_{\Sigma_0} d^3x \sqrt{h}g^{ab}(\phi_1^\ast \nabla_a \phi_2 - \phi_2 \nabla_a \phi_1^\ast)n_b \equiv-\int_{\Sigma_0}d^3x\sqrt{h}j^an_a.\label{eq: innerProd}
\end{align}
It can be shown that $\nabla^aj_a = 0$, which implies that the pseudo inner product is independent of the choice of Cauchy hypersurface. While the pseudo inner product on Minkowski spacetime allows a unique split into positive and negative frequency solutions in flat spacetime, this is no longer true in general for the pseudo inner product \eqref{eq: innerProd} in curved spacetime. Instead, we \emph{choose} a set $\{u_k\}_k$ to denote a complete set of positive frequency orthonormal solutions to the KG equation \eqref{eq: KG curvedST} with respect to this inner product, satisfying $(u_k,u_l)_g=\delta_{kl}$. A basis for the negative frequency solutions is then given by $\{u_k^\ast\}_k$, satisfying $(u_k^\ast,u_l^\ast)_g=-\delta_{kl}$. Note that this choice influences the particle content -- after all, particles are associated to positive frequency modes while anti-particles correspond to solutions with negative frequency. The number of particles is thus, in general, an ambiguous notion in curved spacetime. We can use this basis of positive and negative modes to expand the KG field as
\begin{align}
    \phi(x) = \int d^3k \Big( (u_k,\phi)_g u_k+(-u^\ast_k,\phi)_gu^\ast_k \Big)
\end{align}
and define $a_k \equiv (u_k,\phi)_g$ as well as $a_k^\ast \equiv (-u_k^\ast,\phi)_g$. One can further define a conjugate momentum operator $\Pi(x)$ such that the canonical commutation relations are satisfied. 

We can now quantize the field $\phi(x)$ and the conjugate momentum $\Pi(x)$ by promoting $a_k$ and $a_k^\ast$ to operators $\hat{a}_k$ and $\hat{a}_k^\dagger$. We thus have
\begin{align}
    \hat{\phi}(x) = \int d^3k \left(u_k \hat{a}_k + u_k^\ast \hat{a}_k^\dagger\right).
\end{align}
The operators $\hat{a}_k^\dagger$ and $\hat{a}_k$ have the usual interpretation in terms of creation and annihilation operators and can be used to define the vacuum state
\begin{align}
    \hat{a}_k\ket{0}_g = 0 \hspace{0.5cm} \forall k
\end{align}
as well as the $n$-particle Fock states for fixed $k$
\begin{align}
    \ket{n_k}_g=\frac{1}{\sqrt{n!}}(\hat{a}^\dagger_k)^n\ket{0}_g.
\end{align}

We can apply the same quantization procedure for Minkowski spacetime with modified mass term. One might think that the presence of the spacetime-dependent mass would prevent us from using the usual inner product 
\begin{align}
    (\phi_1,\phi_2)_\eta=i\int_{\Sigma_0} d^3x \eta^{ab}(\phi_1^\ast \partial_a \phi_2 - \phi_2 \partial_a \phi_1^\ast)n_b.\label{eq: innerProdMink}
\end{align}
However, it is straightforward to show that expression \eqref{eq: innerProdMink} is still independent of the Cauchy hypersurface and thus provides a well-defined Hermitian form for Minkowski spacetime. The difference to standard Minkowski spacetime is, however, that the split into positive and negative frequency modes is ambiguous due to the modified dispersion relation that arises as a consequence of the spacetime-dependent mass. Thus, we again make a \emph{choice} of positive frequency modes $\{\tilde{u}_k\}_k$ that are solutions to 
\begin{align}
    \square_\eta \tilde{u}_k(x) = M^2(x) \tilde{u}_k(x)
\end{align}
and satisfy $(\tilde{u}_k,\tilde{u}_l)_\eta=\delta_{kl}$. Together with $\{\tilde{u}_k^\ast\}_k$, satisfying $(\tilde{u}_k^\ast,\tilde{u}_l^\ast)_\eta=-\delta_{kl}$, we can expand the field as follows
\begin{align}
    \tilde{\phi}(x) = \int d^3k \left( (\tilde{u}_k,\tilde{\phi})_\eta \tilde{u}_k+(-\tilde{u}^\ast_k,\tilde{\phi})_\eta\tilde{u}^\ast_k \right).
\end{align}
We define $\tilde{a}_k^\ast \equiv (-\tilde{u}_k^\ast,\tilde{\phi})_\eta$ and $\tilde{a}_k \equiv (\tilde{u}_k,\tilde{\phi})_\eta$ and promote them to creation and annihilation operators $\hat{\tilde{a}}_k^\dagger$ and $\hat{\tilde{a}}_k$. The quantized field is thus
\begin{align}
    \hat{\tilde{\phi}}(x) = \int d^3k \left( \hat{\tilde{a}}_k \tilde{u}_k+\hat{\tilde{a}}_k^\dagger\tilde{u}^\ast_k \right).
\end{align}
Finally, we define the vacuum state as well as the $n$-particle states in modified Minkowski as
\begin{align}
    \hat{\tilde{a}}_k\ket{0}_\eta = 0 \hspace{0.2cm} \forall k \text{ and }
    \ket{n_k}_\eta=\frac{1}{\sqrt{n!}}(\hat{\tilde{a}}^\dagger_k)^n\ket{0}_\eta.
\end{align}

\section{Cosmological particle production} \label{app: cosmological particle production}
Consider the special case of an FLRW spacetime which is asymptotically static in the past and the future, with different scale factors at $\tau\to \pm \infty$. That is, we take a conformally flat metric $g_{ab} = \Omega(x)^{-2} \eta_{ab} = a^2(\tau)\eta_{ab}$ with \begin{align}
    a(\tau) = \begin{cases}
          \cb{a_{in}} \text{ for } \tau \to -\infty,\\
          \cb{a_{out}} \text{ for } \tau \to +\infty.
    \end{cases}
\end{align}
There exists a preferred basis of positive and negative frequency modes $\{ u_k, u_k^\ast \}_k$ and $\{ v_k, v_k^\ast \}_k$ at  $\tau \to -\infty$ and $\tau\to +\infty$ respectively, since there exists a Killing vector field $\partial_\tau$ in these regions with respect to which $\partial_\tau u_k = -i\omega(k)u_k$. This singles out corresponding creation and annihilation operators, denoted by $\hat{a}_k^\dagger$, $\hat{a}_k$ for $\tau \to -\infty$ and $\hat{b}_k^\dagger$, $\hat{b}_k$ for $\tau \to +\infty$ and thus allows us to define unambiguously the number of particles. Because the spacetime under consideration is curved, the operators in the different asymptotic regions are not the same but are related by the Bogoliubov transformations
\begin{align}
    \hat{b}_k &= \sum_{k'}\alpha_{k'k}\hat{a}_k +\beta_{k'k}^\ast \hat{a}^\dagger_{k'},\\
    \hat{b}_k^\dagger &= \sum_{k'}\alpha_{k'k}^\ast \hat{a}_k^\dagger +\beta_{k'k} \hat{a}_{k'}.
\end{align}
Consequently, also the vacuum states $\ket{0_-}$ and $\ket{0_+}$ differ. This leads to particle creation. More concretely, using the number operator $\hat{N}_k^+ = \hat{b}_k^\dagger \hat{b}_k$ defined at $\tau \to +\infty$ to probe the vacuum at $\tau \to -\infty$, the expected
number of particles created in mode $k$ is
\begin{align}
   \expval{\hat{N}_k^+}{0_-}_g = \sum_{k'}|\beta_{k'k}|^2.
\end{align}

\end{document}